\documentclass[showpacs,aps,twocolumn]{revtex4}
\usepackage{graphicx}
\usepackage{amsfonts}
\usepackage{amsmath}
\usepackage{amssymb}
\usepackage{multirow}
\usepackage{braket}
\usepackage[usenames,dvipsnames]{color}

\begin{document}
\bibliographystyle{apsrev}

\title{Entropy and long-range correlations in random symbolic sequences %
}
\author{ S.~S.~Melnik and O.~V.~Usatenko}
\affiliation{ A. Ya. Usikov Institute for Radiophysics and Electronics \\
Ukrainian Academy of Science, 12 Proskura Street, 61805 Kharkov,
Ukraine}
\begin{abstract}

%

The goal of this paper is to develop an estimate for the entropy of
random long-range correlated symbolic sequences with elements
belonging to a finite alphabet. As a plausible model, we use the
high-order additive stationary ergodic Markov chain. Supposing that
the correlations between random elements of the chain are weak we
express the differential entropy of the sequence by means of the
symbolic pair correlation function. We also examine an algorithm for
estimating the differential entropy of \emph{finite} symbolic
sequences. We show that the entropy contains two contributions, the
correlation and fluctuation ones. The obtained analytical results
are used for numerical evaluation of the entropy of written English
texts and DNA nucleotide sequences. The developed theory opens the
way for constructing a more consistent and sophisticated approach to
describe the systems with strong short- and weak long-range
correlations.
\end{abstract}

 \maketitle

\section{Introduction}
Our world is complex, chaotic and correlated. The most peculiar
manifestations of this concept are human and animal communication,
written texts of natural languages, DNA and protein sequences, data
flows in computer networks, stock indexes, solar activity, weather,
etc. For this reason, systems with long-range interactions (and/or
sequences with long-range memory) and natural sequences with
non-trivial information content have been the focus of a large
number of studies in different fields of science for the past
several decades. The unflagging interest in the systems with
correlated fluctuations is also explained by the specific properties
they demonstrate and their prospective applications as a creative
tool for designing the devices and appliances with random components
in their structure (different wave-filters, diffraction gratings,
artificial materials, antennas, converters, delay lines,
etc.~\cite{IzrKrMak}).

Random sequences with \emph{finite number of states} exist as
natural sequences (DNA or natural language texts) or arise as a
result of coarse-grained mapping of the evolution of the chaotic
dynamical system into a string of symbols~\cite{Eren,Lind}. Such
random sequences are the subject of study of the algorithmic
(Kolmogorov-Solomonoff-Chaitin) complexity, artificial intellect,
information theory, compressibility of digital data, statistical
inference problem, computability and have many application aspects
mentioned above.

There are many methods for describing complex dynamical systems and
random sequences connected with them: fractal dimensions,
multi-point probability distribution functions, correlation
functions, and many others. One of the most convenient
characteristics serving to the purpose of studying complex dynamics
is entropy~\cite{Shan,Cover}. Being a measure of the information
content and redundancy in a sequence of data, it is a powerful and
popular tool in examination of complexity phenomena. Among fields of
science where the notion of entropy is of major significance data
compression~\cite{Salomon}, natural language processing~\cite{Mann}
and artificial intelligence~\cite{Wiss} are the most important. The
basic idea of compression is to exploit redundancy in data,
expressed in terms of correlations, and transform this redundancy in
compression algorithm. Recent advances in different fields of
science have hinted at a deep connection between intelligence and
entropy.

A standard method of understanding and describing statistical
properties of a given random sequence of data requires the
estimation of the joint probability function of words occurring for
sufficiently large length $L$ of words. For limited size sequences,
reliable estimations can be achieved only for very small $L$ because
the number $m^L$ (where $m$ is the finite-alphabet length) of
different words of the length $L$ has to be much less than the total
number $M-L$ of words in the whole sequence of the length $M$,
\begin{equation}\label{Strong2}
 m^L\ll M-L\simeq M.
\end{equation}
This is the crucial point because usually the correlation lengths of
natural sequences of interest is of the same order that the length
of sequence. Inequality (\ref{Strong2}) cannot be fulfilled. The
lengths of representative words that could estimate correctly the
probability of words occurring are $4 - 5$ for a real natural text
of the length $10^6$ (written on an alphabet containing $27-30$
letters and symbols) or of order of 20 for a coarse-grained text
represented through a binary sequence. So, long-range correlations
that can exist in the sequences cannot be taken into account in such
a kind of theories.

Here we present a complementary approach, which takes into account
just the long-range correlations. Specifically, we sacrifice the
knowledge of exact statistics of short words and take into account
the weak long-range memory, which can be expressed in terms of the
pair correlation function of symbols and can be found by numerical
analysis of sequence nearly at the same distances as the total
length of sequence.

We use the earlier developed method~\cite{muya} for constructing the
conditional probability function presented by means of pair
correlator, which makes it possible to calculate analytically the
entropy of the sequence. It should be stressed that we suppose that
the correlations are weak but not short. Which kind of memory, long-
or short-range, is more important depends on the intrinsic
correlation properties of the sequence under study.

The scope of the paper is as follows. First, supposing that the
correlations between symbols in the sequence are weak, we 
represent the differential entropy in terms
of the conditional probability function of the Markov chain and
express the entropy as the sum of squares of the pair correlators.
Then we discuss some properties of the results obtained. Next, a
fluctuation contribution to the entropy due to finiteness of random
chains is examined. The application of the developed theory to
literary texts and DNA sequences of nucleotides is considered. In
conclusion, some remarks on directions in which the research can be
progressed are presented.

This work is a generalization of our previous paper \cite{MelUs}
devoted to the binary random sequences. We insistently recommend to
a reader to see it before reading this paper.


\section{Entropy of the additive symbolic Markov chains}

Consider a semi-infinite random stationary ergodic sequence
\begin{equation}\label{ranseq}
 \mathbb{A}= a_{0}, a_{1},a_{2},...
\end{equation}
 of symbols
(letters) $a_{i}$ taken from the finite alphabet
\begin{equation}\label{alph}
 A=\{\alpha^1,\alpha^2,...,\alpha^m\},\,\, a_{i}\in A,\,\, i \in
\mathbb{N}_{+} = \{0,1,2...\}.
\end{equation}
We use the notation $a_i$ to indicate a position of the symbol $a$
in the chain and the notation $\alpha^k$ to stress the value of the
symbol $a\in A$.

We suppose that the symbolic sequence $\mathbb{A}$ is the
\textit{high-order} \emph{Markov
chain}~\cite{Raftery,Ching,Li,Cocho,Seifert}. Such sequences are
also referred to as the multi- or the
$N$-step~\cite{RewUAMM,MUYaG,UYa} Markov's chains. The sequence
$\mathbb{A}$ is the $N$-step Markov's chain if it possesses the
following property: the probability of symbol~$a_i$ to have a
certain value $\alpha^k \in A $ under condition that {\emph{all}
previous symbols are given depends only on $N$ previous symbols,
\begin{eqnarray}\label{def_mark}
&& P(a_i=\alpha^k|\ldots,a_{i-2},a_{i-1})\\[6pt]
&&=P(a_i=\alpha^k|a_{i-N},\ldots,a_{i-2},a_{i-1}).\nonumber
\end{eqnarray}
Sometimes the number $N$ is also referred to as the \emph{order} or
the \emph{memory length} of the Markov chain. Note,
definition~(\ref{def_mark}) is valid for $i\geqslant N$; for $i<N$
we should use the well known conditions of compatibility for the
conditional probability functions of lower order~\cite{shir}.

To estimate the differential entropy of stationary sequence
$\mathbb{A}$ of symbols $a_{i}$ one could use the Shannon definition
\cite{Shan} for entropy per block of length $L$,
\begin{eqnarray} \label{entro_block}
H_{L}=-\sum_{a_{1},...,a_{L} \in A} P(a_{1}^{L})\log_{2}
P(a_{1}^{L}).
\end{eqnarray}
Here $P(a_{1}^{L}) =P(a_{1},\ldots,a_{L})$ is the probability to
find $L$-word $a_{1}^{L}$ in the sequence; hereafter we use the more
concise notation $a_{i-N}^{i-1}$ for $N$-word $a_{i-N},...,a_{i-1}$.
The differential entropy, or the entropy per symbol, is given by
\begin{eqnarray} \label{entro_diff}
h_{L}=H_{L+1} - H_{L}.
\end{eqnarray}
This quantity specifies the degree of uncertainty of $(L+1)$th
symbol occurring and measures the average information per symbol if
the correlations of $(L+1)$th symbol with preceding $L$ symbols are
taken into account. The differential entropy $h_L$ can be
represented in terms of the conditional probability function
$P(a_{L+1}|a_{1}^{L})$, 
\begin{eqnarray} \label{Entro_Bin}
h_L=\!\!\sum_{a_{1},...,a_{L} \in A}\!\! P(a_{1}^{L})
h(a_{L+1}|a_{1}^{L}) = \overline{ h(a_{L+1}|a_{1}^{L})} ,
\end{eqnarray}
where $h(a_{L+1}|a_{1}^{L})$ is the amount of information contained
in the $(L+1)$th symbol of the sequence conditioned on $L$ previous
symbols,
\begin{eqnarray}
   h(a_{L+1}|a_{1}^{L}) = - \!\!\sum_{a_{L+1} \in A}\!\!
P(a_{L+1}|a_{1}^{L})\log_2 P(a_{L+1}|a_{1}^{L}).
    \label{siL}
\end{eqnarray}
The source entropy (or Shannon entropy) is the differential entropy
at the asymptotic limit, $h=\lim_{L \rightarrow \infty}h_{L}$. This
quantity  measures the average information per symbol if {\it all}
correlations, in the statistical sense, are taken into account, cf.
with \cite{Grass}, Eq.~(3).

Due to the ergodicity of stationary sequence $\mathbb{A}$, the
average value of any function $f(a_{r_1},a_{r_1+r_2},\ldots ,
a_{r_1+\ldots+r_{s}})$ of $s$ arguments defined on the set $A$ of
symbols is statistical (arithmetic, Cesaro's) average over the
chain,
\begin{eqnarray}\label{epsilon-av}
&& \overline{f}\,(a_{r_1},\ldots ,
a_{r_1+\ldots+r_{s}})\\[6pt]
&&=\lim_{M\to\infty}\frac{1}{M} \sum_{i=0}^{M-1}f(a_{i+r_1},\ldots ,
a_{i+r_1+\ldots+r_{s}}). \nonumber
\end{eqnarray}

Stationarity together with decay of correlations,
\emph{$C_{\alpha,\beta}(r\to\infty)=0$}, see below
definition~(\ref{Dev_eq1}), leads, according to the Slutsky
sufficient conditions~\cite{Slu}, to mean-ergodicity. This latter
property is very useful in numerical calculations since the
averaging can be done over the length of the sequence and the
ensemble averaging can be avoided. Therefore, in our numerical as
well as analytical calculations we always apply averaging over the
length of the sequence as it is implied in Eq.~(\ref{Entro_Bin}).

If the sequence, statistical properties of which we would like to
analyze, is given, the conditional probability function (CPF) of
$N$th order can be found by a standard method (written below for
subscript $i=N+1$)
\begin{equation}\label{soglas}
P(a_{N+1}=\alpha^k|a_{1},\ldots,a_{N})=\frac{
P(a_{1},\ldots,a_{N},\alpha^k) } { P(a_{1},\ldots,a_{N})},
\end{equation}
where $P(a_{1},\ldots,a_{N},\alpha^k)$ and $P(a_{1},\ldots,a_{N})$
are the probabilities  of the $(N+1)$-subsequence
$a_{1},\ldots,a_{N},\alpha^k$ and $N$-subsequence
$a_{1},\ldots,a_{N}$ occurring, respectively.

The Markov chain with CPF of general form Eq.~(\ref{def_mark}) is
not convenient (compliant) to solve concrete problems. For this
reason we introduce a simplification for the CPF. Specifically, we
suppose that the symbolic Markov chain under consideration is
\textit{additive}, i.e. its conditional probability is a linear
function of random variables $a_k,\,\, k=i-N,...,i-1$,
\begin{equation}\label{Dev_eq2}
P(a_i=\alpha|a_{i-N}^{i-1})=p_{\alpha}+\sum_{r=1}^N  \sum_{\beta \in
A} F_{\alpha \beta}(r)[\delta(a_{i-r},\beta)-p_{\beta}],
\end{equation}
where $p_{\alpha}$ is the relative number of symbols $\alpha$ in the
chain, or their probabilities of occurring,
\begin{equation}\label{a-av}
p_{\alpha}=\overline{\delta(a_i , \alpha)}.
\end{equation}
Here $\delta(.,.)$ is the Kronecker delta-symbol, playing the role
of the characteristic function of the random variable $a_i$ and
converting symbols to numbers. Hereafter, we often drop the
superscript $k$ from $\alpha^k$ to simplify the notations.

The additivity means that the previous symbols $a_{i-N}^{i-1}$ exert
an independent effect on the probability of the symbol $a_i=\alpha$
occurring. The first term in the right-hand side of
Eq.~(\ref{Dev_eq2}) is responsible for correct reproduction of
statistical properties of uncorrelated sequences, the second one
takes into account, and produces under generation, correlations
among symbols of the random sequence. The conditional probability
function in form~(\ref{Dev_eq2}) can reproduce correctly the binary
(pair, two-point) correlations in the chain. Higher-order
correlators and all correlation properties of higher orders are not
independent anymore. We cannot control them and reproduce correctly
by means of the memory function $F_{\alpha \beta}(r)$ because the
latter is completely determined by the pair correlation function,
see below Eq.~(\ref{Dev_eq4}).

The additive Markov chains are, in some sense, analogous to the
chains described by autoregressive
models~\cite{Raftery,Chakravarthy}. In Appendix A some suggestions
on the form of Eq.~(\ref{Dev_eq2}) and its properties are presented.

There is a rather simple relation between the memory  function
$F_{\alpha \beta}(r)$ and the pair \emph{symbolic} correlation
function of the additive Markov chain. The two-point symbolic
correlation function is defined as
\begin{equation}\label{Dev_eq1}
C_{\alpha \beta}(r)\!= \!\overline{\big[\delta(a_i,
\alpha)-p_{\alpha} \big]\!\big[ \delta(a_{i+r},
\beta)-p_{\beta}\big]}, \,\,\alpha,\beta\!\in \!\!A.
\end{equation}
This function possesses the following properties:
\begin{eqnarray}
&& C_{\alpha \beta}(r)=C_{\beta \alpha}(-r),   \\ [1pt]
&&\sum_{\alpha \in A} C_{\alpha \beta}(r)=\sum_{\beta \in A}
C_{\alpha \beta}(r)=0.\nonumber
\end{eqnarray}

Let us suppose that there exists a one-to-one correspondence
$a_i\leftrightarrow\varepsilon_i$ between the letters of symbolic
sequence $\mathbb{A}$ and the numbers of numeric sequence. Then, the
ordinary ``numeric'' correlation function
\begin{equation}\label{Dev_eq8}
C_{\varepsilon}(r)=\overline{(\varepsilon_i-\bar{\varepsilon})
(\varepsilon_{i+r}-\bar{\varepsilon})}
\end{equation}
of the sequence of $\varepsilon_i$ can be expressed by means of
symbolic correlator

\begin{equation}\label{Def Rel}
C_{\varepsilon}(r)=\sum_{\alpha,\beta\in
A}\varepsilon^\alpha\varepsilon^\beta C_{\alpha \beta}(r).
\end{equation}
Here $\varepsilon^\alpha$ is the numeric value of the random
variable $\varepsilon$ corresponding to the symbol $\alpha$,
$\sum_{\alpha \in A}$ means the summation over all possible letters
of the alphabet $A$.

There were suggested two methods for finding $F_{\alpha \beta}(r)$
of a sequence with a known pair correlation function. The first
one~\cite{muya} is based on the minimization of the ``distance''
between the conditional probability function, containing the
sought-for memory function, and the given sequence $\mathbb{A}$ of
symbols with a known correlation function,
\begin{equation}\label{Dev_eq40}
Dist = \overline{[\delta(a_{i},\alpha) -
P(a_i=\alpha|a_{i-N}^{i-1})]^2}.
\end{equation}

 For any values of $\alpha,\beta \in A$ and $r \geqslant 1$ the minimization
equation with respect to $F_{\alpha \beta}(r)$ yields the
relationship between the correlation and memory functions,
\begin{equation}\label{Dev_eq4}
C_{\alpha \beta}(r)=\sum_{r'=1}^{N} \sum_{\gamma \in A}C_{\alpha
\gamma}(r-r') F_{\beta \gamma}(r').
\end{equation}
The second method for deriving Eq.~(\ref{Dev_eq4}) is a completely
probabilistic straightforward calculation analogous to that used
in~\cite{MUYaG}.

Equation~(\ref{Dev_eq4}), despite its simplicity, can be
analytically solved only in some particular cases: for one- or
two-step chains, the Markov chain with a step-wise memory function
and so on. To avoid the various difficulties in its solving we
suppose that correlations in the sequence are weak (in amplitude,
but not in length). In order to formulate this condition we
introduce the \emph{normalized} symbolic correlation function
defined by
\begin{equation}\label{Def K}
K_{\alpha \beta}(r)=\frac{C_{\alpha \beta}(r)}{C_{\alpha \beta}(0)},
\quad C_{\alpha \beta}(0)=p_{\alpha}\delta(\alpha,\beta)-p_{\alpha}
p_{\beta}.
\end{equation}
We can obtain an approximate solution
for the memory function in the form of the series
\begin{equation}\label{Series}
F_{\alpha \beta}(r)=K_{\beta \alpha}(r) + \sum_{r'\neq r}^N
\sum_{\gamma \in A}
 K_{\gamma \alpha}(r-r')K_{\beta \gamma}(r')+ ...
\end{equation}
if we suppose the all components of the normalized correlation
function with $r \neq 0$ are small with respect to $K_{\alpha
\beta}(0)=1$.

Equation~(\ref{Dev_eq2}) for the conditional probability function in
the first approximation with respect to the small parameters
$|K_{\alpha \beta}(r)| \ll 1, \, r\neq 0$ after neglecting the
second term in~Eq.(\ref{Series}) takes the form
\begin{equation}\label{Approx CP}
P(a_i=\alpha|a_{i-N}^{i-1})\simeq p_{\alpha}+\sum_{r=1}^N
\sum_{\beta \in A} K_{\beta
\alpha}(r)[\delta(a_{i-r},\beta)-p_{\beta}].
\end{equation}

This formula provides a tool for constructing weak correlated
sequences with a given pair correlation function~\cite{muya}. Note
that $i$-independence of the function $P(a_i=\alpha|a_{i-N}^{i-1})$
provides homogeneity and stationarity of the sequence under
consideration; and finiteness of $N$ together with the strict
inequalities
\begin{equation}\label{ergo_m}
 0 < \!P(a_{i+N}\!=\alpha|a_{i}^{i+N-1})\! < 1, \, i \in \mathbb{N}_{+} = \{0,1,2...\}
\end{equation}
provides, according to the Markov theorem (see, e.g.,
Ref.~\cite{shir}), ergodicity of the sequence.

The conditional probability $P(a_i=\alpha|a_{i-L}^{i-1})$ for a word
of length $L < N$ can be obtained in the first approximation in the
weak correlation parameter $\Delta_{\alpha}(L)$ from
Eqs.~\eqref{Dev_eq2} and \eqref{Approx CP} by means of a routine
probabilistic reasoning presented in Appendix B,
\begin{eqnarray} \label{p_i(L)}
P(a_i&=&\alpha|a_{i-L}^{i-1}) = p_{\alpha} + \Delta_{\alpha}(L), \\
\Delta_{\alpha}(L) &=& \sum_{r=1}^L \sum_{\beta \in A}
K_{\beta\alpha}(r)[\delta(a_{i-r},\beta)-p_{\beta}]. \nonumber
\end{eqnarray}

Taking into account the weakness of correlations,
\begin{eqnarray}\label{weakness of cor}
|\Delta_a(L)| \ll 1,
\end{eqnarray}
we expand Eq.~\eqref{siL} in Taylor series up to the second order in
$\Delta_{\alpha}(L)$, $h(a_{L+1}|a_{1}^{L}) = h_0 + (\partial
h/\partial p_{\alpha})\Delta_{\alpha}(L) + (1/2)(\partial^2
h/\partial p_{\alpha}^2)\Delta^2_{\alpha}(L)$, where the derivatives
are taken at the point $P(a_i=\alpha|a_{i-L}^{i-1}) = p_{\alpha}$
and $h_0$ is the entropy of uncorrelated sequence,
\begin{eqnarray}
h_0=-\sum_{\alpha \in A}p_{\alpha}\log_{2}(p_{\alpha}).
\end{eqnarray}
Then, the differential entropy of the sequence in line with $
\overline{\Delta_{\alpha}(L)} =0$ takes the form 
\begin{equation}\label{Entro_Markov2}
h_L= \left\{\begin{array}{l} h_{L<N}= h_0 -
\dfrac{1}{2\ln2}\sum_{r=1}^L \sum_{\alpha \in A}
\frac{\overline{\Delta^2_{\alpha}(L)}}{p_{\alpha}}, \\[8pt]
 h_{L>N}=h_{L=N}.
\end{array}
\right.
\end{equation}

If the length of block exceeds the memory length, $L>N$, the
conditional probability $P(a_{i}=\alpha|a_{i-L}^{i-1})$ depends only
on $N$ previous symbols, see Eq.~(\ref{def_mark}). Then, it is easy
to show from~\eqref{Entro_Bin} that the differential entropy remains
constant at $ L \geqslant N$. Thus, the second line in
Eq.~\eqref{Entro_Markov2} is consistent with the first line because
in the first approximation in the weak correlations the parameter
$\Delta_\alpha(L)$ vanishes at $L>N$ together with the correlation
function.  The final expression, the main analytical result of the
paper, for the differential entropy of a stationary ergodic weakly
correlated random sequence is
\begin{equation}\label{EntroMain}
h_L =  h_0 - \frac{1}{2\ln2}\sum_{r=1}^L \sum_{\alpha, \beta \in A}
\frac{C_{\alpha\beta}^2(r)}{p_{\alpha} p_{\beta}}.
\end{equation}

In order to obtain this equation we used Eq.~\eqref{p_i(L)} and
replaced the term $C_{\alpha \beta}(r'-r)$ with $C_{\alpha
\beta}(0)\delta(r,r')$ when calculating the summation.

\section{Discussion}

It follows from Eq.~(\ref{EntroMain}) that the additional correction
to the entropy $h_0$ of the uncorrelated sequence is negative. This
is the anticipated result -- the correlations decrease the entropy.
The conclusion is not sensitive to the sign of correlations:
persistent correlations, $K>0$, describing an ``attraction'' of the
symbols of the same kind, and anti-persistent correlations, $K<0$,
corresponding to a ``repulsion'' between the same symbols, provide
the corrections of the same negative sign. If the correlation
function is constant at $1 \leqslant r \leqslant N$, the entropy is
a linear decreasing function of the argument $L$ up to the point
$r=N$.

Equation~\eqref{EntroMain} takes more simple form for a binary,
$m=2$, chain of symbols, which can be also considered as a numeric
chain of random variables $a_i$ with the alphabet of symbols-numbers
$A = \{0;1\}$. Let $p_1=\bar{a}$, $p_0=1-\bar{a}$. In order to
calculate $h_L$ we should calculate four symbolic correlation
functions:
\begin{eqnarray}
&& C_{11}(r)=\overline{\delta(a_i,1) \delta(a_{i+r},1)} - \bar{a}^2,
\\ [3pt]\nonumber && C_{00}(r)=\overline{\delta(a_i,0) \delta(a_{i+r},0)}
- (1 - \bar{a})^2 ,\\ [3pt] \nonumber &&
C_{01}(r)=\overline{\delta(a_i,0) \delta(a_{i+r},1)} - (1 -
\bar{a})\bar{a} ,\\[3pt] \nonumber && C_{10}(r)=\overline{\delta(a_i,1)
\delta(a_{i+r},0)} - \bar{a}(1 - \bar{a}). \nonumber
\end{eqnarray}
Taking into account that $\delta(a_i,1) = a_i$, è $\delta(a_i,0) = 1
- a_i$, we obtain
\begin{eqnarray}
&& C_{11}(r)=C_{00}(r)= C(r),
 \\[2pt]
&& C_{01}(r)= C_{10}(r) =  - C(r). \nonumber
\end{eqnarray}
Here $C(r)$ is the ordinary numeric correlator
\begin{equation}\label{def_cor0}
C(r)=\overline{(a_i-\bar{a})(a_{i+r}-\bar{a})}.
\end{equation}
After simple algebra, we get
\begin{equation}
h_L =   h_0 - \frac{1}{2\ln2} \sum_{r=1}^L K^2(r),
\end{equation}
where $K(r)$ is the normalized pair correlation function of the
binary sequence $K(r)=C(r)/C(0)$, the result obtained earlier in
Ref.~\cite{MelUs}.

\section{Finite random sequences}

The relative numbers $p_{\alpha}$ of symbols in the chain,
correlation functions and other statistical characteristics of
random sequences are deterministic quantities only in the limit of
their infinite lengths. It is a direct consequence of the law of
large numbers. If the sequence length $M$ is finite, the set of
numbers $a_1^M$ cannot be considered anymore as ergodic sequence. In
order to restore its status we have to introduce the \emph{ensemble}
of finite sequences $\{a_1^M\}_p, p \in \mathbb{N} =0,1,2,...$. Yet,
we would like to
retain the right to examine \emph{finite} sequences 
by using a single finite chain. So, for a finite chain we have to
replace definition~(\ref{Dev_eq1}) of the correlation function by
the following one,
\begin{eqnarray}\label{CorFin}
C_{\alpha \beta,M}(r)\!\!& =& \!\!\frac{1}{M-r}
\!\!\!\!\sum_{i=0}^{M-r-1}\!\!\!\! \big[\delta(a_i,
\alpha)-p_{\alpha} \big]\big[ \delta(a_{i+r},
\beta)-p_{\beta}\big], \nonumber \\
  p_{\alpha}&=&\frac{1}{M}\sum_{i=0}^{M-1}\delta(a_i,\alpha),
\end{eqnarray}
which coincides with Eq.~(\ref{Dev_eq1}) in the limit
$M\rightarrow\infty$. Now the correlation functions and $p_{\alpha}$
are random quantities, which depend on the particular realization of
the sequence $a_1^M$. Fluctuations of these random quantities can
contribute to the entropy of finite random chains even if the
correlations in the random sequence are absent. It is well known
that the order of relative fluctuations of additive random quantity
(as, e.g. the correlation function Eq.~(\ref{CorFin})) is
$1/\sqrt{M}$.

Below we give more rigorous justification of this explanation and
show its applicability to our case. Let us present the correlation
function $C_M(r)$ as the sum of two components,
\begin{equation}\label{CorrelSquar}
C_{\alpha \beta,\,M}(r)= C_{\alpha \beta}(r)+C_{\alpha
\beta,\,f}(r),\,\,r\geqslant 1,
\end{equation}
where the first summand $C_{\alpha
\beta}(r)=\lim_{M\rightarrow\infty} C_{\alpha \beta,\,M}(r)$ is the
correlation function determined by Eq.~(\ref {CorFin}) (in the limit
$M\rightarrow\infty$) obtained by averaging over the sequence with
respect to index $i$, enumerating the elements $a_{i}$ of sequence
$\mathbb{A}$; and the second one, $C_{\alpha \beta,\,f}(r)$, is a
fluctuation--dependent contribution. Function $C_{\alpha \beta}(r)$
can be also presented as the ensemble average $C_{\alpha
\beta}(r)=\langle C_{\alpha \beta,\,M}(r) \rangle$ due to the
ergodicity of the (infinite) sequence.

Now we can find a relationship between variances of $C_{\alpha
\beta,\,M}(r)$ and $C_{\alpha \beta,\,f}(r)$. Taking into account
Eq.~(\ref {CorrelSquar}) and the properties $\langle C_{\alpha
\beta,\,f}(r) \rangle =0$ at $r\neq 0$ and $C_{\alpha
\beta}(r)=\langle C_{\alpha \beta,\,M}(r) \rangle$  we have
\begin{equation}\label{CorrelSquar1}
\langle C_{\alpha \beta,\,M}^2(r) \rangle = C_{\alpha \beta}^2(r) +
\langle C_{\alpha \beta,\,f}^2(r) \rangle, \,\,r\geqslant 1.
\end{equation}

The correlation function $C_{\alpha \beta}(r)$ vanishes when $r$
exceeds the correlation length $R_c$, $r\gg R_c$. It makes possible
to find the asymptotical value of $C_{\alpha \beta,\,f}^2(r)$
\begin{eqnarray}
 \langle C_{\alpha \beta,\,f}^2(r)\rangle_{|r\gg R_c}& \cong
\langle C_{\alpha \beta,\,M}^2(r) \rangle & =  \\ [4pt]
\frac{1}{(M-r)^2}\langle\!\!\sum_{i,j=0}^{M-r-1}\!\!&\!\!\big[\delta(a_i,
\alpha)-p_{\alpha} \big]&\!\!\!\!\big[ \delta(a_{i+r},
\beta)-p_{\beta}\big]\nonumber \\  \times &\!\!\big[\delta(a_j,
\alpha)-p_{\alpha} \big]&\!\!\!\!\big[ \delta(a_{j+r},
\beta)-p_{\beta}\big]\rangle. \nonumber \label{Fluct}
\end{eqnarray}

Neglecting the correlations between elements $a_i$ and taking into
account that the terms with $i=j$ give the main contribution to the
result,
\begin{eqnarray}\label{Fluct2}
 \langle & \sum_{i,j=0}^{M-r-1}&\big[\delta(a_i, \alpha)-p_{\alpha}
\big]\big[ \delta(a_{i+r}, \beta)-p_{\beta}\big] \nonumber \\
&\times& \big[\delta(a_j, \alpha)-p_{\alpha} \big]\big[
\delta(a_{j+r}, \beta)-p_{\beta}\big]\rangle \nonumber \\
 &\cong&\!\!\!\!\!\!\!\!\!\!\!\!
\sum_{i=0}^{M-r-1}\langle\big[\delta(a_i, \alpha)-p_{\alpha} \big]^2
\rangle\langle\big[ \delta(a_{i+r}, \beta)-p_{\beta}\big]^2\rangle \nonumber \\
&=&\!\!\!\!\!\!\!\!(M-r)\,\,C_{\alpha \alpha, f}(0)C_{\beta \beta,
f}(0).
\end{eqnarray}
we obtain, after neglecting $r$ in the term $M-r$, the averaged
fluctuation-dependent contribution to the squared correlation
function
\begin{eqnarray}\label{FluctFin}
\langle C_{\alpha \beta,\,f}^2(r)\rangle &\simeq&
\frac{1}{M}C_{\alpha \alpha, f}(0)C_{\beta \beta, f}(0), \\[6pt]
C_{\alpha \beta}(0)&=&p_{\alpha}\delta(\alpha,\beta)-p_{\alpha}
p_{\beta}.\nonumber
\end{eqnarray}

Note that Eq.~(\ref{FluctFin}) is obtained by means of averaging
over the ensemble of chains. This is the shortest way to get the
desired result. At the same time, for numerical simulations we have
only used  the averaging over the chain as is seen from
Eq.~(\ref{CorFin}), where the summation over sites $i$ of the chain
plays the role of averaging.

Note also that the different symbols $a_i$ in Eq.~(\ref{Fluct}) are
correlated. It is possible to show by direct evaluation of
$C_{\alpha \beta,\,f}^2(r)$ with CPF~(\ref{Approx CP}) that the
contribution of their correlations to $\langle C_{\alpha
\beta,\,f}^2(r) \rangle$ is of order of $\Delta/M^2\ll 1/M$.

Equation~(\ref{EntroMain}), containing $C_{\alpha \beta}(r)$, is
only valid  for the infinite chain. In reality, we always work with
sequences of finite length and can calculate $C_{\alpha
\beta,\,M}(r)$, which contains the fluctuating part. To improve
result~(\ref{EntroMain}) we have to subtract the fluctuating part of
entropy, proportional to $\sum_{r=1}^L \langle C_{\alpha
\beta,\,f}^2(r) \rangle$,  from Eq.~(\ref{EntroMain}).
Thus, Eqs.~(\ref{CorrelSquar1}) and~(\ref{FluctFin}) yield the
differential entropy of the \emph{finite} weakly correlated
(approximately ergodic, $R_c\ll M$) random sequences
\begin{equation}\label{EntroFin}
h_L=  h_0 - \frac{1}{2\ln2}\left[\sum_{r=1}^L \sum_{\alpha, \beta
\in A} \frac{C_{\alpha\beta, \,M}^2(r)}{p_{\alpha}
p_{\beta}}-(m-1)^2  \frac{L}{M}\right].
\end{equation}

It is clear that in the limit $M\rightarrow\infty$ this function
transforms into Eq.~(\ref{EntroMain}).  The last term in RHS of
Eq.~(\ref{EntroFin}) describes the linearly decreasing fluctuation
correction of the entropy. For the binary chain, $m=2$, we get the
result obtained earlier in~\cite{MelUs}.

The squared correlation function $C_{\alpha\beta, \,M}^2(r)$ is
normally a decreasing function of $r$, whereas the function
$C_{\alpha \beta,\,f}^2(r)$ is nearly constant (see
Eq.~(\ref{FluctFin}) for $r\ll M$). Hence, the terms $\sum_{r=1}^L
\sum_{\alpha, \beta \in A} C_{\alpha\beta, \,M}^2(r)/p_{\alpha}
p_{\beta}$ and $(m-1)^2 L/M$ being concave and linear functions,
respectively, describe the competitive contributions to the entropy.
It is not possible to analyze all particular cases of their
relationship. Therefore we indicate here the most interesting ones
taking in mind monotonically decreasing correlation functions. An
example of such a type of function is $C(r)=a/r^{b}, \, a>0, \,b
> 0 $.
\begin{figure}[h!]
\center\includegraphics[width=0.44\textwidth]{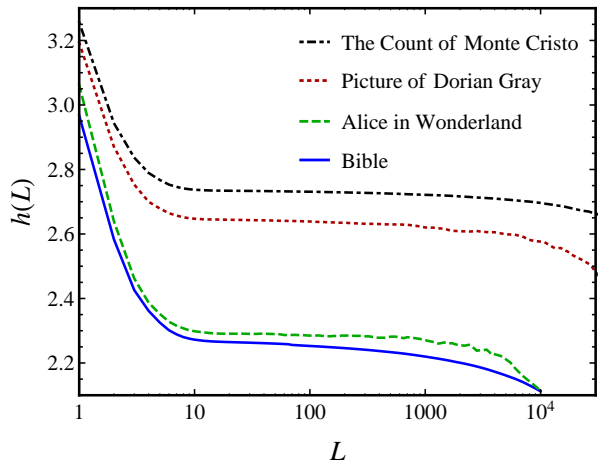}
\caption{(Color online) The differential entropy of the literature
works (indicated in the legend near the curves) vs the length of
words in $L$-axis log scale. The curves correspond to the direct
evaluations of Eq.~(\ref{EntroMain}) with fluctuation correction. }
\label{Alice_eng}
\end{figure}
If the correlations are extremely small and compared with the
inverse length $M$ of the sequence, $\sum_{\alpha, \beta \in A}
C_{\alpha\beta, \,M}^2(1)/p_{\alpha} p_{\beta} \sim 1/M$, the
fluctuating part of the entropy exceeds the correlation part almost
for all values of $L>1$.

When the correlations are more strong, $\sum_{\alpha, \beta \in A}
C_{\alpha\beta, \,M}^2(1)/p_{\alpha} p_{\beta}> 1/M$, there is at
least one point where the contribution of fluctuation and
correlation parts of the entropy are equal. For monotonically
decreasing function $\sum_{\alpha, \beta \in A} C_{\alpha\beta,
\,M}^2(r)/p_{\alpha} p_{\beta}$ there is only one such point.
Comparing the functions in square brackets in Eq.~(\ref{EntroFin})
we find that they are equal at some $L = R_{s}$, which hereafter
will be referred to as a stationarity length. If $L \ll R_s$, the
fluctuations of the correlation function are negligibly small with
respect to its magnitude, hence for these $L$-words the finite
sequence may be considered as the quasi-stationary one. At $ L \sim
R_{s}$ the fluctuations are of the same order as the genuine
correlation function contribution, $\sum_{\alpha, \beta \in A}
C_{\alpha\beta, \,M}^2(r)/p_{\alpha} p_{\beta}$. Here we have to
take into account the fluctuation correction due to the finiteness
of the random chain. At $ L > R_{s}$ the fluctuation contribution
exceeds the correlation one and Eq.~(\ref{EntroFin}) loses any
sense.

The other important parameter of the random sequence is the memory
length $N$. If the length $N$ is less than $R_{s}$, we have no
difficulties to calculate the entropy of the finite sequence, which
can be considered as quasi-stationary. If the memory length exceeds
the stationarity length, $R_{s} \lesssim N$, we should take into
account the fluctuation correction to the entropy.

\section{Applications to natural and DNA texts}\label{Appl}

The purpose of this section is to illustrate applicability of the
developed theory to some concrete sequences naturally arising in
biology and linguistics.

In order to evaluate the differential entropy of literature works we
calculate the probabilities $p_{\alpha}$ of each letter occurring in
the simplified text and  symbolic correlation functions
$C_{\alpha\beta, \,M}(r)$. The simplification (some sort of
coarse-graining) consists in replacing all the upper-case letters
with the lower-case ones and neglecting all punctuation marks except
blanks. Hence, we use the alphabet of 27 letters. The result for
calculating the differential entropy with the use of
Eq.~(\ref{EntroFin}) is shown in Fig.~\ref{Alice_eng}. The entropy
per one letter $h(0)$ (not shown in the picture) is $4\pm0.1$. It is
evident that the difference between the one-letter-entropy, in the
case of the letters equipartition $\log_2 27 \approx 4.75$, and
$4\pm0.1$ is due to the non-equipartition distribution of letters in
the texts.

As we mentioned, the correlation length can be determined as the
length where the entropy takes on a constant value. At first glance,
the value of $R_c$ is of order of $9-11$. But after this point we
observe a nearly linear small decrease of entropy extended over
$2-3$ decades. Probably, this phenomenon could be explained by small
power-low correlation observed and discussed in Ref.~\cite{MUYaG}.

Application of the developed theory to nucleotide sequences of DNA
molecules is shown in Fig.~\ref{fig1DNA}. In order to evaluate the
entropy of the \emph{Homo sapiens} chromosome Y, locus NW 001842422
\cite{ncbi},  we calculate the probabilities $p_{\alpha}$ of each
nucleotide occurring in the sequence and 9 different symbolic
correlation functions $C_{\alpha\beta, \,M}(r)$.

\begin{figure}[h!]
\center\includegraphics[width=0.47\textwidth]{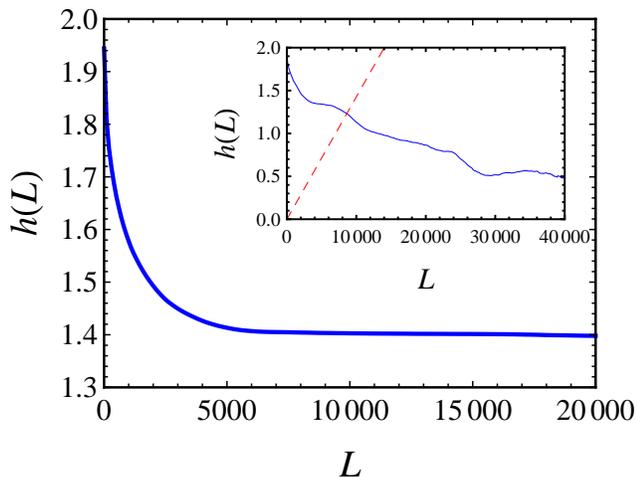}
\caption{(Color online) The differential entropy of \emph{Homo
sapiens} chromosome Y, locus NW 001842422 \cite{ncbi}, of length
$M\simeq3.9\times 10^6$ vs length $L$ with the fluctuation
correction. The curve is constructed by using Eq.~(\ref{EntroMain}).
The inset demonstrates the differential entropy of \emph{Homo
sapiens} chromosome Y, locus NW 001842451, of length $M\simeq
4.5\times 10^4$. The straight dashed line is fluctuation correction
$9L/2\ln2\, M$ due to finiteness of chain.} \label{fig1DNA}
\end{figure}

It is clearly seen that the entropy in the interval $7\times
10^3<L<2\times 10^4$ takes on the constant value, $h_{L}\simeq
1.41$. It means that for $L>7\times 10^3$  all binary correlations,
in the statistical sense, are taken into account. In other words,
the correlation length of the \emph{Homo sapiens} chromosome Y is of
the order of $10^4$. This length $R_c$ is much grater than
correlation length $R_c\approx 10$ observed for natural written
texts.

In the inset the differential entropy of \emph{Homo sapiens}
chromosome Y, locus NW 001842451, is shown. Here we cannot see a
constant asymptotical region, which would be
an evidence for the existence of stationarity 
and finiteness of the correlation length. We suppose that the locus
is not well described by our theory at long distances due to the
relatively short length of sequence. The dashed line in the figure
is the fluctuation correction of the differential entropy. This
correction should be small with respect to the correlation
contribution in the region of reliability of the result. Thus, only
for $L<10^3$ the result can be considered as a plausible.
\begin{figure}[h!]
\center\includegraphics[width=0.44\textwidth]{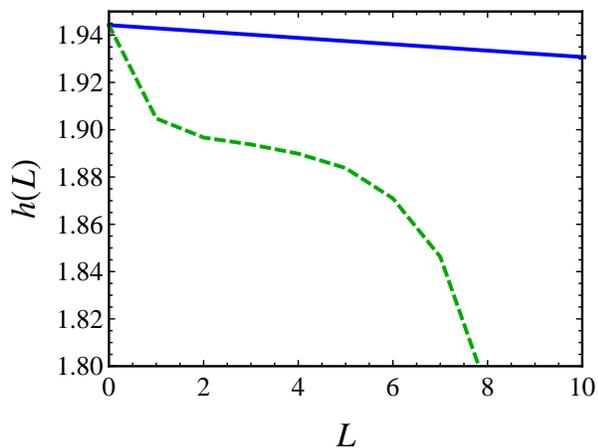}
\caption{(Color online) Comparison of differential entropies
calculated by estimation of block occurring Eq.~(\ref{Estim})
(bottom curve) and the result of Eq.~(\ref{EntroFin}) (top curve)
for the \emph{Homo sapiens} chromosome Y, locus NW 001842422. }
\label{figDNA_block}
\end{figure}

It is interesting to compare our results with those obtained by
estimation of block entropy Eq.~(\ref{entro_block}) where the
probabilities of words occurring are calculated with standard
likelihood estimate
\begin{equation}\label{Estim}
P(a_{1}^{L})=\frac{n(a_{1}^{L})}{M-L+1}.
\end{equation}
Here $n(a_{1}^{L})$ is the number of occurrences of the word
$a_{1}^{L}$ in the sequence of the length $M$. In our paper
\cite{MelUs} it was shown that there is a good agreement between two
approaches for the coarse-grained (binary) DNA sequence of R3
chromosome of \emph{Drosophila melanogaster} of length
$M\simeq2.7\times 10^7$ for $L\lesssim 5-6$ units. For four-valued
sequence (composed by adenine, guanine, cytosine, thymine) we cannot
make a similar conclusion studying the differential entropy of the
\emph{Homo sapiens} chromosome Y, locus NW 001842422, shown in
Fig.~\ref{figDNA_block}. It is clear that at small $L$ strong
short-range correlations or the exact statistics of the short words
are more important than that which we took into account --- the
simple pair correlations.

It is difficult to come to an unambiguous conclusion, which factor,
the finiteness of the chain and violation of Eq.~(\ref{Strong2}) or
the strength of correlations, is more important for the discrepancy
between the two theories and between the two studied sequences.

\section{Conclusion and perspectives}

 (i) The main result of the paper, the differential entropy of the
stationary ergodic weakly correlated random sequence $\mathbb{A}$
with elements belonging to the finite alphabet is given by
Eq.~(\ref{EntroMain}). The other important point of the work is the
calculation of the fluctuation contribution to the entropy due to
the finiteness of random chains, the last term in
Eq.~(\ref{EntroFin}).

(ii) In order to obtain Eq.~(\ref{EntroMain}) we used an assumptions
that the random sequence of symbols is the high-order Markov chain.
Nevertheless, the final result contains only the correlation
function and does not contain the conditional probability function
of the Markov chain. This allows us to suppose that
result~(\ref{EntroMain}) and the region of its applicability is
wider than the assumptions under which it is obtained.

(iii) To obtain Eq.~(\ref{EntroMain}) we supposed that the
correlations in the random chain are weak. It is not a very severe
restriction. Many examples of such kind of systems described by
means of the pair correlator are given in Ref. \cite{IzrKrMak}. The
randomly chosen example of DNA sequences and the literary texts
support this conclusion. The strongly correlated systems, which is
opposed to weakly correlated chains, are nearly deterministic. For
their description we need completely different approach. Their study
is beyond the scope of this paper.

(iv) Equation~(\ref{EntroMain}) can be considered as an expansion of
the entropy in series with respect to the small parameter $\Delta$,
where the entropy $h_0$ of the non-correlated sequence is the zero
approximation. Alternatively, for the zero approximation we can use
the exactly solvable model of the $N$-step Markov chain with the
conditional probability function of words occurring taken in the
form of the step-wise function \cite{UYa}. Another way to choose the
zero approximation can be based on CPF obtained from probability of
the block occurring Eq.~(\ref{entro_block}). Consequently, the
developed theory opens the way to construct a more consistent and
sophisticated approach describing the systems with strong
short-range and weak long-range correlations.

(v) Our consideration can be generalized to the Markov chain with
the infinite memory length $N$. In this case we should impose the
condition of the decreasing rate of the correlation function and the
conditional probability function at $N\rightarrow \infty$.

\appendix

\section{}

The conditional probability function of the \emph{binary additive}
Markov chain of random variables $a_i\in \{0, 1\}$, the probability
of symbol~$a_i$ to have a value $1$ under the condition that $N$
previous symbols $a_{i-N}^{i-1}$ are given, is of the following
form~\cite{muya,RewUAMM},
\begin{equation} \label{prob1}
 P(a_i=1|a_{i-N}^{i-1})) = \bar{a} + \sum_{r=1}^N F(r)(a_{i-r} -
\bar{a}).
\end{equation}
Analogously for $P(0|.)$,
\begin{eqnarray} \label{prob0}
P(a_i&=&0|a_{i-N}^{i-1}) =1-
P(1|a_{i-N}^{i-1})\nonumber \\
&=&1-\bar{a}-\sum_{r=1}^N F(r)(a_{i-r} - \bar{a}).
\end{eqnarray}
This two expressions are not symmetric with respect to the change $0
\leftrightarrows 1$ of generated symbol $a_i$. Let us show that
Eqs.~(\ref{prob1}) and~(\ref{prob0}) can be presented in the
symmetric form
\begin{equation}\label{prob2}
P(a_i=\alpha|a_{i-N}^{i-1})\!=\! p_{\alpha}+\!\!\sum_{r=1}^N \!
\sum_{\beta \in \{0,1\}}\!\!\! F_{\alpha
\beta}(r)[\delta(a_{i-r},\beta)-p_{\beta}].
\end{equation}

Taking into account the definitions $p_1=\bar{a}$, $p_0=1-\bar{a}$,
using the  evident equalities $\delta(a_{i-r},0)=1-a_{i-r}$,
$\delta(a_{i-r},1)=a_{i-r}$ and putting $ F_{1 1}(r)- F_{1 0}(r)
=F_{00}(r)- F_{01}(r)= F(r)$  we easily obtain Eqs.~(\ref{prob1})
and (\ref{prob0}). We should replace $\alpha,\beta \in \{0,1\}$ in
Eq.~(\ref{prob2}) by $\alpha,\beta \in A $ to obtain
Eq.~(\ref{Dev_eq2}).

Note, there is no one-to-one correspondence between the memory
function $F_{\alpha \beta}(r)$ and the conditional probability
function $P(a_i=\alpha|a_{i-N}^{i-1})$. Indeed, it is easy to see
that, in view of Eqs.~(\ref{Dev_eq2}) and~(\ref{a-av}), the
renormalized memory function $F'_{\alpha \beta}(r)=F_{\alpha
\beta}(r)+\varphi_{\alpha} (r)$ provides the same conditional
probability as $F_{\alpha \beta}(r)$.

\section{}

Here we prove Eq.~(\ref{p_i(L)}) using Eqs.~(\ref{Dev_eq2}) and
(\ref{Approx CP}) as a starting point. It follows from definition
(\ref{soglas}) of the conditional probability function
\begin{equation}\label{App1}
P(a_i=a|W)=\frac{P(W,a)} { P(W)}, \quad W = a_{i-N+1}^{i-1}.
\end{equation}
Adding symbol $a_{i-N}=b$ to the string $(W,a)$ we have
\begin{equation}\label{App2}
P(a_i=a|W)=\frac{\sum_{b \in A}P(b,W,a)} { P(W)}.
\end{equation}

Replacing here the probabilities $P(b,W,a)$ by the CPF
$P(a_i=a|b,W)$ from the equation similar to that of
Eq.~(\ref{App1}),
\begin{equation}\label{App2'}
P(a_i=a|b,W)=\frac{P(b,W,a)} { P(b,W)},
\end{equation}
we obtain after some algebraic manipulations
\begin{eqnarray}
 &&P(a_i=a|W)= p_a + \sum_{r=1}^{N-1} \sum_{b \in A}
F_{ab}(r)[\delta(a_{i-r},b)-p_b]
\nonumber\\[6pt]
&&+\frac{1}{P(W)}\sum_{c\in A}F_{ac}(N)\sum_{b\in A}
 P(b,W)\left[ \delta(b,c)-p_c\right].
\label{App3}
\end{eqnarray}

The 3-rd term containing summation over $b$ is of the form
\begin{equation}\label{App3'}
 P(c,W)(1-p_c)-P(\overline{c},W)p_c,
 \end{equation}
where the symbol $\overline{c}$ stands for an event NOT-$c$. It is
intuitively clear that in the zero approximation in $\Delta$ (i.e.,
for uncorrelated sequence) this term equals zero. In the next
approximation this term is of order of $\Delta$. These two
statements can be verified by using the condition of compatibility
for the Chapman-Kolmogorov equation (see, for example,
Ref.~\cite{gar}),
\begin{equation}\label{App4}
P(a_{i-N+1}^{i})=\sum_{a_{i-N}\in A}P(a_{i-N}^{i-1})
P_{N}(a_i|a_{i-N}^{i-1}).
\end{equation}
Hence, we have to neglect the third term in the right-hand side of
Eq.~(\ref{App3}) because it is of the second order in $\Delta$. So,
Eq.~(\ref{p_i(L)}) is proven for $L=N-1$. By induction, the equation
can be written for arbitrary $L$.

\begin{acknowledgments} We are grateful for the helpful and fruitful
discussions with G.~M.~Pritula, S.~S.~Apostolov, and Z.~A.~Maizelis.
\end{acknowledgments}

\end{document}